\documentclass[aps,prl,twocolumn,superscriptaddress,floatfix,amsmath,amssymb]{revtex4-2}
\usepackage{subfigure}
\usepackage[english]{babel}
\usepackage{graphicx}
\usepackage{dcolumn}
\usepackage{bm}
\usepackage{verbatim}
\usepackage{mathrsfs}
\usepackage{natbib}
\usepackage{cancel}
\usepackage{color}
\usepackage[normalem]{ulem}

\def\slashchar#1{\setbox0=\hbox{$#1$} 
\dimen0=\wd0 
\setbox1=\hbox{/} \dimen1=\wd1 
\ifdim\dimen0>\dimen1 
\rlap{\hbox to \dimen0{\hfil/\hfil}} 
#1 
\else 
\rlap{\hbox to \dimen1{\hfil$#1$\hfil}} 
/ 
\fi}

\begin{document}
\title{Reentrant phase transitions in Einstein-Maxwell-scalar black holes
}
\author{Dumitru Astefanesei}
\affiliation{Pontificia Universidad Cat\'olica de Valpara\'iso, Instituto de F\'isica, Av. Brasil 2950, Valpara\'iso, Chile}
\author{Paulina Cabrera}
\affiliation{Universidad T\'ecnica Federico Santa Mar\'ia, Departamento de F\'isica, Av. Espa\~na 1680, Valpara\'iso, Chile}
\author{Robert B. Mann}
\affiliation{Department of Physics and Astronomy, University of Waterloo, Waterloo, Ontario, N2L 3G1, Canada}
\affiliation{Perimeter Institute for Theoretical Physics, 31 Caroline St N, Waterloo, ON, N2L 2Y5, Canada}
\author{Ra\'ul Rojas}
\affiliation{Pontificia Universidad Cat\'olica de Valpara\'iso, Instituto de F\'isica, Av. Brasil 2950, Valpara\'iso, Chile}

\begin{abstract}
We consider how scalar fields affect the thermodynamic behavior of charged anti-de Sitter (AdS) black holes. We specifically investigate a class of (3+1)-dimensional exact hairy charged AdS black hole	solutions to Einstein-Maxwell-scalar gravity, whose stable ground state and finite horizon area in the zero temperature limit make it of particular interest. We find that the reverse isoperimetric inequality is satisfied for this class and that there exists an intermediate range of the charge that admits reentrant phase behavior, the first example of this type of phase behavior in (3 + 1) dimensions in a consistent theory.
\end{abstract}
\date{October 2021}
\maketitle

\noindent\emph{Introduction ---} There is considerable ongoing interest in how scalar fields affect gravitational systems, particularly in asymptotically anti-de Sitter (AdS) spacetimes.  Scalars can condense to form `boson stars'  (smooth horizonless compact objects in AdS) \cite{Astefanesei:2003qy,Astefanesei:2003rw,Buchel:2013uba}
and can also yield black hole spacetimes with a single Killing vector
\cite{Dias:2011at,Stotyn:2011ns,Stotyn:2012ap, Henderson:2014dwa,Brihaye:2014nba,Herdeiro:2015kha}.
Hairy black hole solutions play an important role  in the context of AdS-CFT duality \cite{Maldacena:1997re} in understanding various properties of field theories with a holographic dual, e.g. \cite{Nastase:2017cxp,Ammon:2015}.  Surprisingly, it was  recently shown that the self-interaction of the scalar field is relevant for the stability of black holes in asymptotically flat spacetime \cite{Astefanesei:2019mds,Astefanesei:2019qsg,Astefanesei:2020xvn}. 

Consequently, the thermodynamic behavior of hairy black holes is also of considerable importance.  It is of particular interest to understand how the presence of scalars modifies the phase behavior of charged black holes \cite{Kubiznak:2012wp,Gunasekaran:2012dq}, something well understood in the context of black hole chemistry \cite{Kubiznak:2014zwa}.
This approach extends   the framework of black hole thermodynamics  to allow  for a `dynamical pressure' and its conjugate volume \cite{Kubiznak:2016qmn}. The AdS cosmological constant is taken to be a thermodynamic variable \cite{Henneaux:1985tv, Creighton:1995au}  interpreted as thermodynamic pressure \cite{Kastor:2010gq, Kastor:2011qp}.  Its conjugate volume  is conjectured to satisfy a relation called the Reverse Isoperimetric Inequality \cite{Cvetic:2010jb}  whose violation may be associated with a new kind of
instability for black holes  \cite{Johnson:2019mdp,Cong:2019bud}.  Once embedded in string/M theory,  changing the cosmological constant is equivalent to a geometrical process, namely one with a  variation of the volume (modulus) of  the external sphere.

In this paper, we show that scalar hair can drastically modify known phase behavior.  Specifically, we
present the first example of reentrant phase transitions in four spacetime dimensions and in a consistent theory describing a class of exact hairy charged AdS black hole solutions to  Einstein-Maxwell-scalar gravity. These are of particular interest because they have finite horizon area in the zero temperature limit \cite{Anabalon:2013sra}, whereas without the dilaton potential, these   solutions are singular.  This is essential for a correct definition of  the canonical ensemble when the charge of black hole is fixed \cite{Chamblin:1999tk}. Furthermore, the dilaton potential of the theory corresponds to an extended supergravity model with dyonic Fayet-Iliopoulos terms and so has a well defined (stable) ground state \cite{Anabalon:2017yhv, Anabalon:2020pez}.  Such solutions  are important in considering quantum phase transitions \cite{Sachdev:2019bjn, Anabalon:2021smx} and merit further investigation since  the dilaton potential modifies the `AdS box', suggesting unexpected new features and possible insight into  distinct holographic phases of matter \cite{Charmousis:2010zz}.

We show here that this class \cite{Anabalon:2013sra, Anabalon:2012ta} of charged hairy black holes does indeed exhibit an interesting range of phase behavior depending on their charge.   In particular, there is an intermediate range of charge for
 which, as pressure increases, these black holes go from exhibiting no distinguishable phases, to a sequence of reentrant phase transitions, to a standard first-order Van der Waals phase transition, to a critical point, to a state where there are again no distinguishable phases. 
 
\noindent\emph{Extended thermodynamics  and the reverse isoperimetric inequality ---}
We consider the Einstein-Maxwell-dilaton theory with the action
\begin{equation}
I=\frac{1}{2\kappa}
\int_{\mathcal{M}}
{d^4x\sqrt{-g}\left[R-e^{\sqrt{3}\phi}F^2
	-\frac{(\partial\phi)^{2}}{2}-U\right]}
\label{sqrt3action}
\end{equation}
where $\kappa=8\pi$ in the unit system where $G=c=1$. The self-interaction of the scalar field is
\begin{align}
U&=\frac{\alpha}{8}\left[ \sinh(\sqrt{3}\phi)
+9\sinh\left(\frac{\phi}{\sqrt{3}}\right)
 -{4\sqrt{3}}\phi\cosh\left(\frac{\phi}{\sqrt{3}}\right)\right]\nonumber\\
 &\qquad + 2\Lambda \cosh\left(\frac{\phi}{\sqrt{3}}\right)
\label{pot}
\end{align}
where $\alpha$ is an arbitrary parameter and $\Lambda$ is the cosmological constant. The exact static spherically symmetric solution can be put in the following form \cite{Anabalon:2013sra}
\begin{align}
ds^2&=\Omega\left[-fdt^2
+\eta^2f^{-1}dx^2
+d\theta^2+\sin^2\theta d\varphi^2\right] \\ \label{gauge}
\Omega&=\frac{4x}{\eta^2(x^2-1)^2}\,, \quad A= \left(-\frac{q}{2x^2}+\frac{q}{2x_+^2}\right)dt,
\end{align}
and $\phi(x)=\sqrt{3}\ln(x)$ for the scalar field, where $x$ is a dimensionless radial coordinate and $\eta$ and $q$ are constants of integration. Without loss of generality, we can assume $\eta\geq 0$. As in Ref. \cite{Chamblin:1999hg}, the gauge potential has been fixed such that $A_t(x_+)=0$, with $x_+$ being the horizon coordinate satisfying $f(x_+)=0$, where
\begin{align}
\label{metrics}
f&=-\frac{\Lambda}{3} +\frac{\alpha}{4}\left[\ln(x)+\frac{x}{\eta^2\Omega} -\sqrt{\frac{x}{\eta^2\Omega}}\right]+\frac{x}{\Omega} -\frac{2q^2}{\eta\sqrt{x\Omega^3}}
\end{align}
The solution is known to admit two different branches, but we consider the positive branch for which $x>1$. The boundary is located at $x\rightarrow 1$, where the scalar field vanishes, and the singularity is at $x=\infty$. 
We require the conformal factor $\Omega(x)\to r^2$, with $r$ being the Schwarzschild radial coordinate, in order to obtain the canonical AdS boundary. We obtain the Ashtekar-Magnon-Das mass \cite{Ashtekar:1984zz,Ashtekar:1999jx},
\begin{equation}
\label{AMD}
M=\frac{3\eta^2(2q^2-1)-\alpha}{6\eta^3}
\end{equation}
The physical electric charge obtained by the Gauss law is related to the parameters of the solution by $Q=q/\eta$. The Hawking temperature and entropy can be computed as
\begin{equation}
T=-\frac{1}{4\pi\eta}f'(x_+)\,, \qquad S=\frac{\mathcal{A}}{4}=\pi\Omega(x_+)
\end{equation}

Now, by considering the cosmological constant to be a variable corresponding to a spacetime pressure, we have the generalized first law $dM=TdS+\Phi dQ+VdP$, where
where
\begin{equation}
P\equiv -\frac{\Lambda}{8\pi}\,,\quad   V\equiv\left(\frac{\partial{M}}{\partial{P}}\right)_{Q,S}
=\frac{8\pi}{3\eta^3}\frac{3x_{+}^2+1}{(x_{+}^2-1)^3}
\end{equation}
are the respective pressure and its conjugate thermodynamic volume. Furthermore, the Reverse Isoperimetric Inequality \cite{Cvetic:2010jb}, $\mathcal{R} \geq 1$, is satisfied
\begin{equation}
\mathcal{R}\equiv\left(
\frac{3V}{4\pi}\right)^{\frac{1}{3}} 
\left(\frac{4\pi}{\mathcal{A}}\right)^{\frac{1}{2}} =\frac{2^{\frac{1}{3}}(3x_+^2+1)^{\frac{1}{3}}}{2\sqrt{x_+}}\geq 1
\end{equation}
 
\noindent\emph{Criticality in canonical ensemble ---}
In what follows, we shall consider  the thermodynamics in the extended phase space by keeping the electric charge fixed. The thermodynamic behavior of the black hole is governed by the thermodynamic potential $\mathcal{F}=M-TS$ obtained from the on-shell action in the Euclidean section \cite{Anabalon:2015xvl},
\begin{equation}
\label{quantumsr}
\mathcal{F}=\beta^{-1}I^E =\beta\left(\frac{\alpha}{12\eta^3}-\frac{{\eta}Q^2}{2{x_+^2}} +\frac{1}{2\eta}\frac{x_+^2+1}{x_+^2-1}\right)
\end{equation}

In what follows, we shall rescale all variables in terms of 
the fixed positive constant $\alpha$
\begin{align}
\label{rescaled}
&\eta\rightarrow\frac{\eta}{\sqrt{\alpha}} \,, \quad P\rightarrow\frac{P}{\alpha}\,, \quad V\rightarrow\alpha^{\frac{3}{2}}V \,, \quad T\rightarrow\frac{T}{\sqrt{\alpha}}, \nonumber\\
&S\rightarrow \alpha S \,, \quad Q\rightarrow\sqrt{\alpha}Q \,, \quad M\rightarrow\sqrt{\alpha}M,
\end{align}
such that all thermodynamic quantities are dimensionless; hence 
 $\alpha$ will not further appear explicitly.

\begin{figure}[t!]
\centering
{\includegraphics[width=5.2cm]{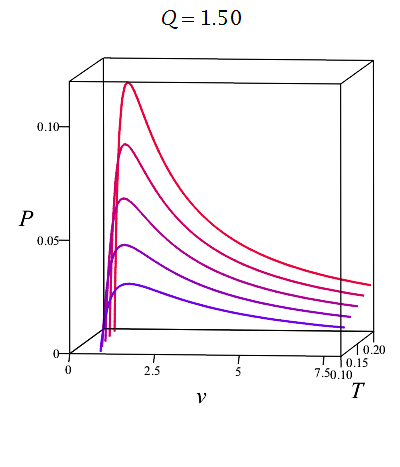}} {\includegraphics[width=5.2cm]{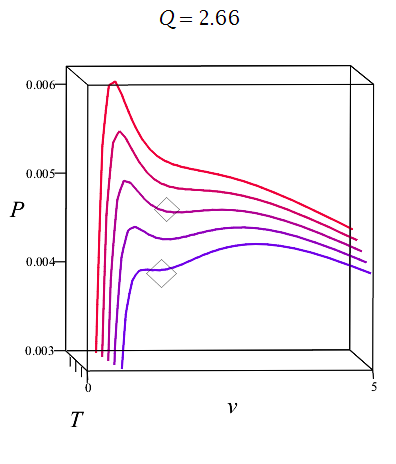}}
{\includegraphics[width=5.2cm]{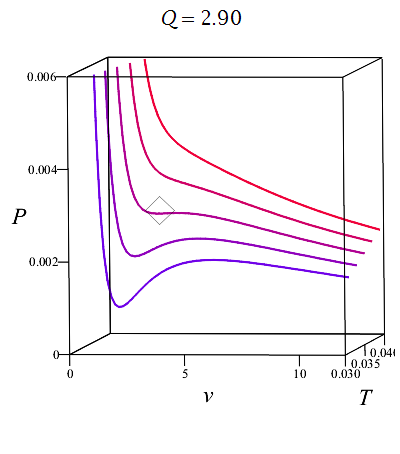}}
\caption{\small Equation of state for $Q=1.50$, $Q=2.65$ and for $Q=2.90$, respectively. Double criticality is observed within $Q_{min}\approx 2.622<Q<Q_0\approx 2.712$.}
\label{PVT}
\end{figure}

In this setup, we start by studying the equation of state $P=P(T,v)$, where $v\equiv 3V/2S$ is the specific volume, at fixed values of the electric charge $Q$. The equation of state cannot be analytically obtained, but it can be expressed parametrically as $T=T(x_+,P,Q)$ and $v=v(x_+,P,Q)$, and it is depicted in Fig. \ref{PVT}
As shown, the phase behavior changes as $Q$ takes different values. There are three branches each with its characteristic behavior. For small values of the electric charge, $Q<Q_{min}\approx 2.622$, the different isotherms exhibit a similar behavior pattern: they start from a minimum value of $v$ at $P=0$, and then, as $v$ increases, $P$ reaches a maximum, after which it decreases as $v$ continues to increase. For intermediate values, $Q_{min}\leq Q\leq Q_0\approx 2.712$, the behavior is dependent on the temperature, as shown in the second plot in Fig. \ref{PVT}. Within this interval, two critical points appear. The critical point at higher pressure corresponds to that of a standard Van der Waals phase transition and, for temperatures just below this, there is a standard first-order large/small black hole phase transition. The critical point at lower pressure corresponds to a new feature whose nature will be clarified later on in this section. For lower temperatures there are no longer two distinct phases. Finally, for $Q>Q_0$, the behavior becomes of the same type as the Reissner-Nordstr\"om-AdS (RN-AdS) solution, with only one critical point. We shall see that one of the critical points in the region $Q_{min}<Q<Q_0$ of ``double criticality" corresponds to a transition between two unstable phases and thus is of no physical interest.

\begin{figure}[h!]
\centering
{\includegraphics[width=8 cm]{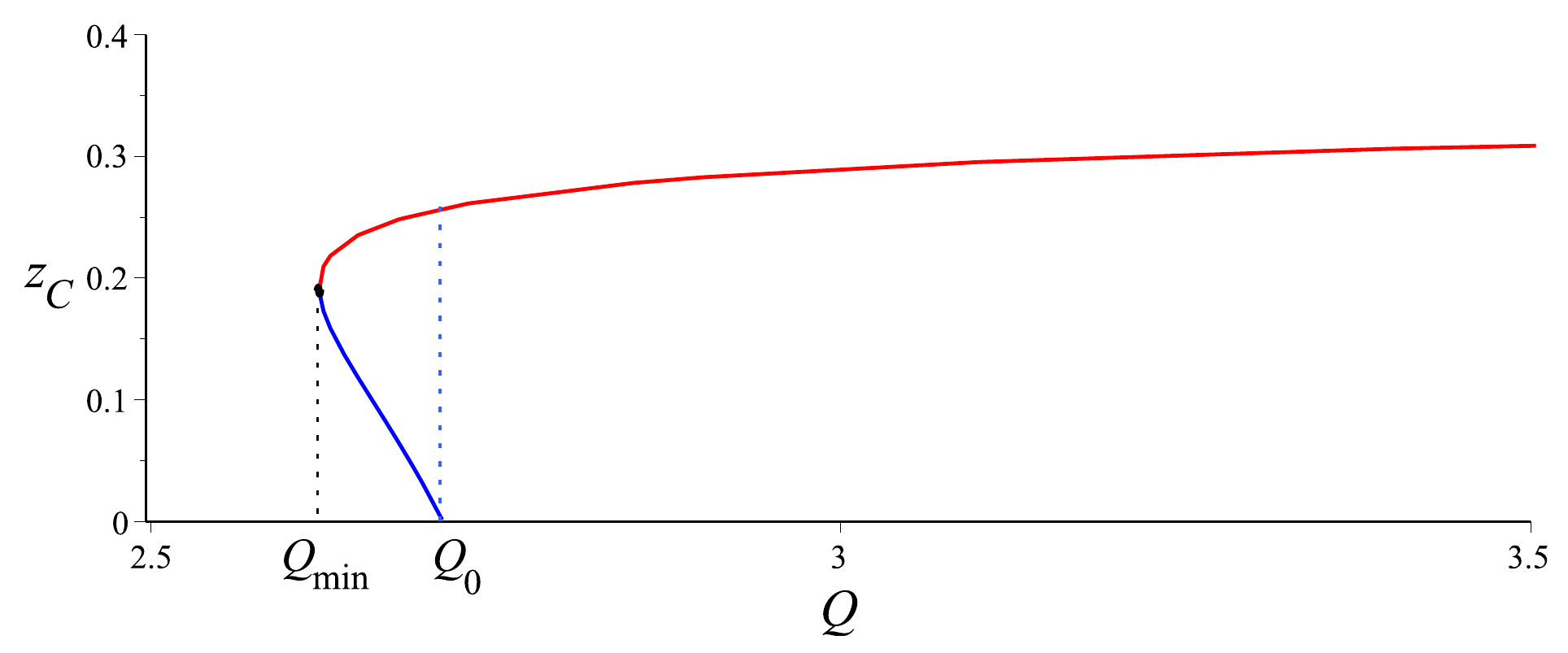}}
{\includegraphics[width=8 cm]{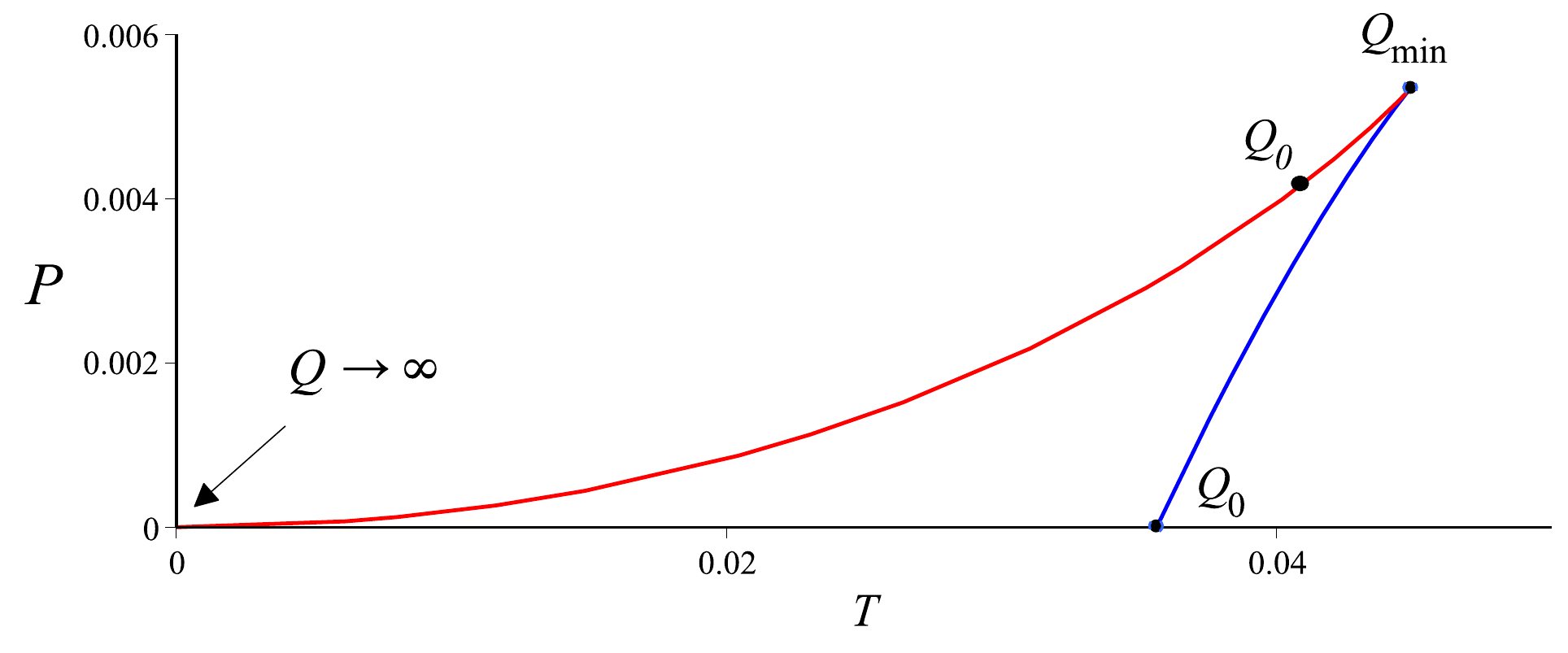}}
\caption{\small Top: $z_C\equiv P_cv_c/T_c$ vs $Q$. Note that $z_C<3/8$, with $z_C\rightarrow 3/8$ as $Q\rightarrow\infty$. {Bottom:} trajectory of critical points in the $P-T$ plane, parametrized by $Q$.}
\label{PT}
\end{figure}

The critical points, denoted as $(P_c,v_c,T_c)$, satisfy
\begin{equation}
\left(\frac{\partial P}{\partial v}\right)_{T_C,Q}=0,
\qquad
\left(\frac{\partial^2 P}{\partial v^2}\right)_{T_C,Q}=0
\end{equation}
The single critical point for $Q>Q_0$ is characterized by a critical compressibility factor $z_C\equiv P_cv_c/T_c$ that, in the limit $Q\rightarrow\infty$, approaches  $3/8$, as expected. This RN-AdS-like critical point extends to $Q=Q_{min}$, where both critical points meet. This is shown in the first plot in Fig.~\ref{PT}, where we observe the double criticality within $Q_{min}\leq Q\leq Q_0$. 
As the electric charge continuously changes, the critical points trace a trajectory in the $P-T$ plane, as depicted in the second plot in Fig. \ref{PT}. We also notice that both critical points meet at $T_c\approx 0.0448$ and $P_c\approx 0.0053$ when $Q\rightarrow Q_{min}$.

\begin{figure}[t!]
\centering
\includegraphics[width=4.1 cm]{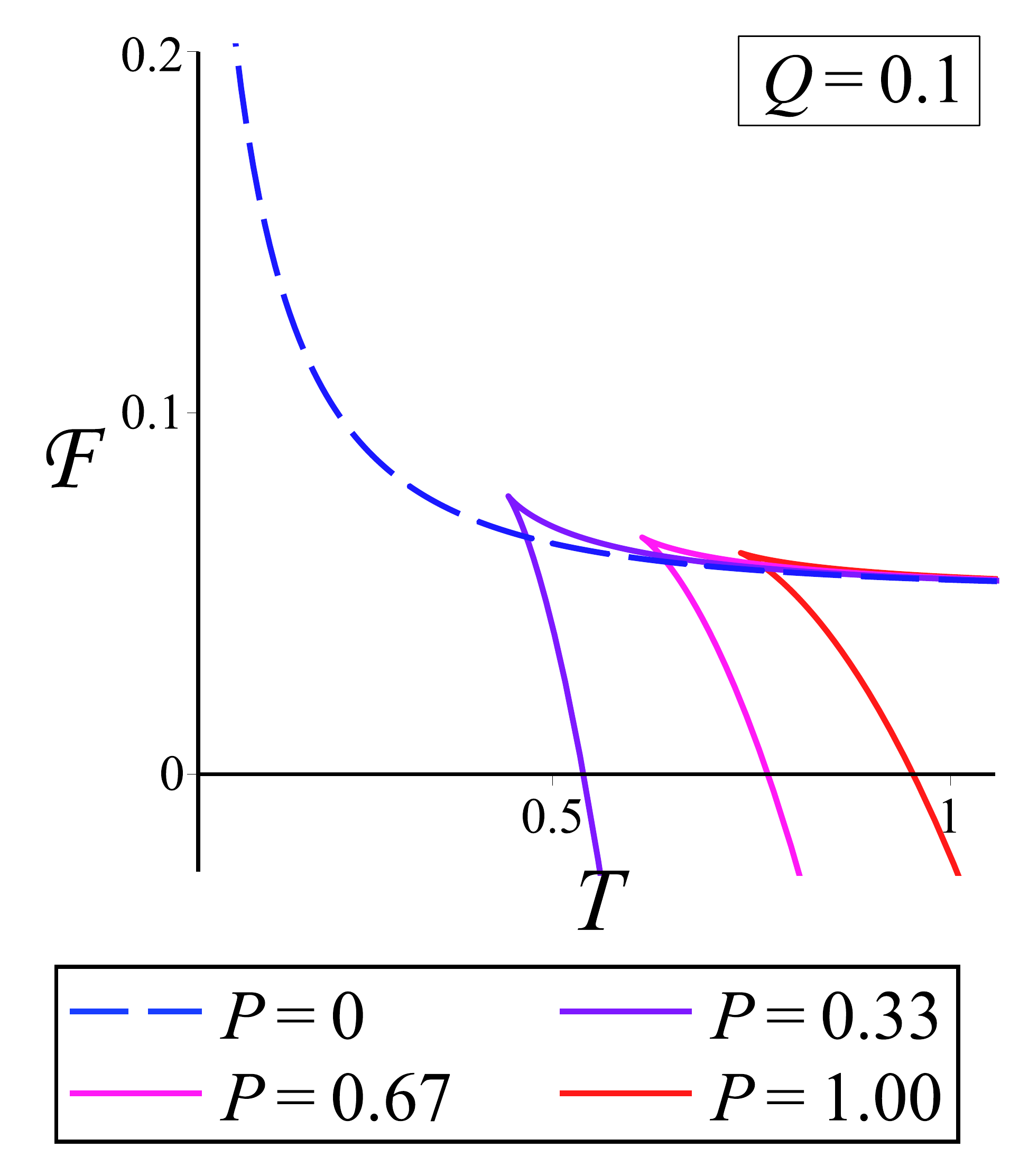}
\includegraphics[width=4.1 cm]{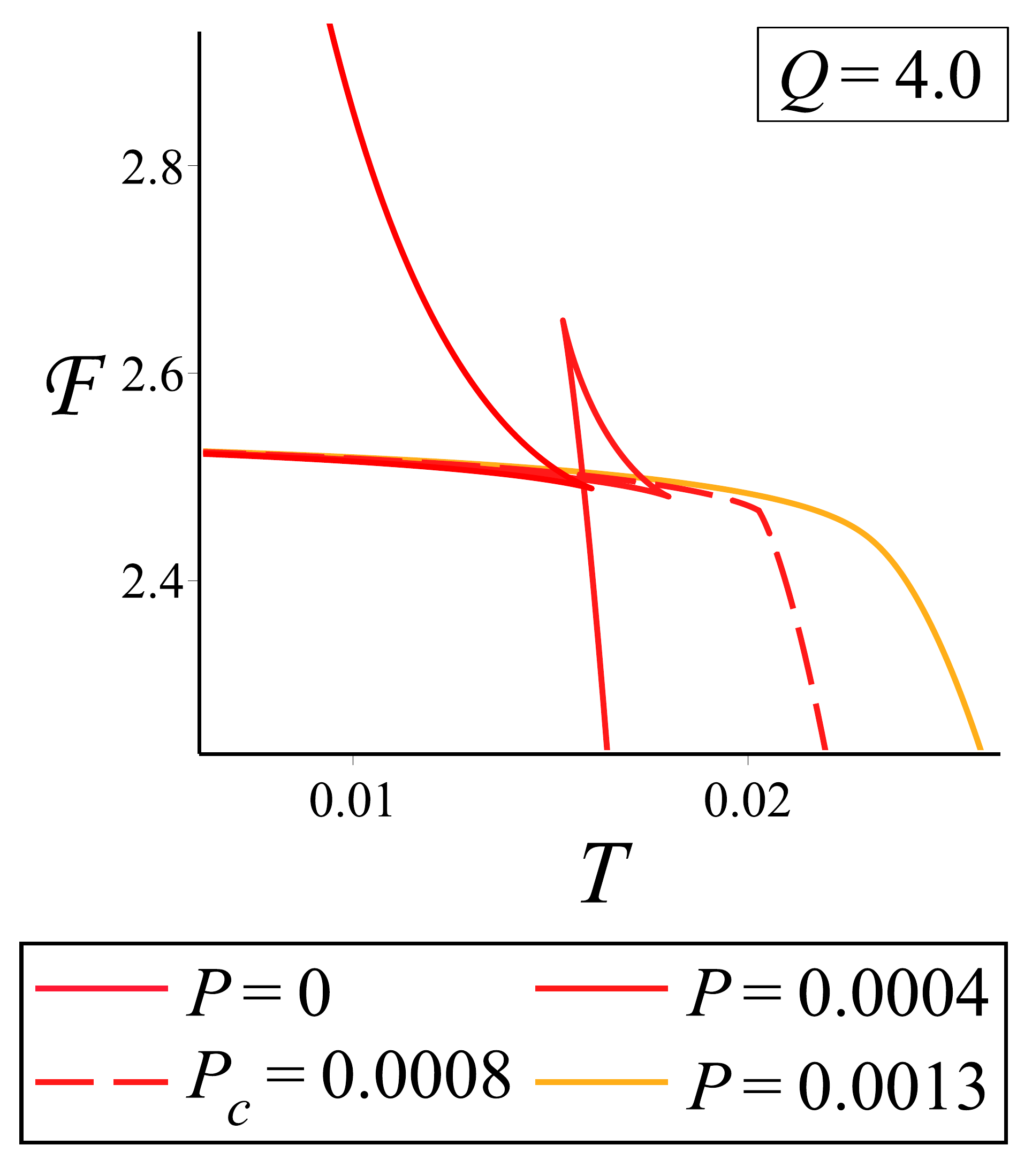}
\caption{\small Left hand side: $\mathcal{F}$ vs $T$ for $Q=0.10$ ($<Q_{min}$) for which there is no critical behavior. {Right hand side:} Here we consider $Q=4$ ( $>Q_0$) for which the critical behavior is similar than the RN-AdS black hole.}
\label{extremal}
\end{figure}

We now study the phase behavior using the thermodynamic potential (\ref{quantumsr}). For $Q<Q_{min}$, no critical behavior takes place (see the first plot in Fig. \ref{extremal}) and, for $Q>Q_0$, the behavior is quite similar to the RN-AdS black hole, with one critical point and a small-to-large first order phase transition, as shown in the second plot in Fig. \ref{extremal}. 
The cusps at the left in the cases $Q<Q_0$ and also $Q_{min}<Q<Q_0$ are indicative of the fact that these black holes have no  inner horizon and hence no extremal limit.

\begin{figure}[t!]
\centering
\subfigure[]
{\includegraphics[scale=0.17]{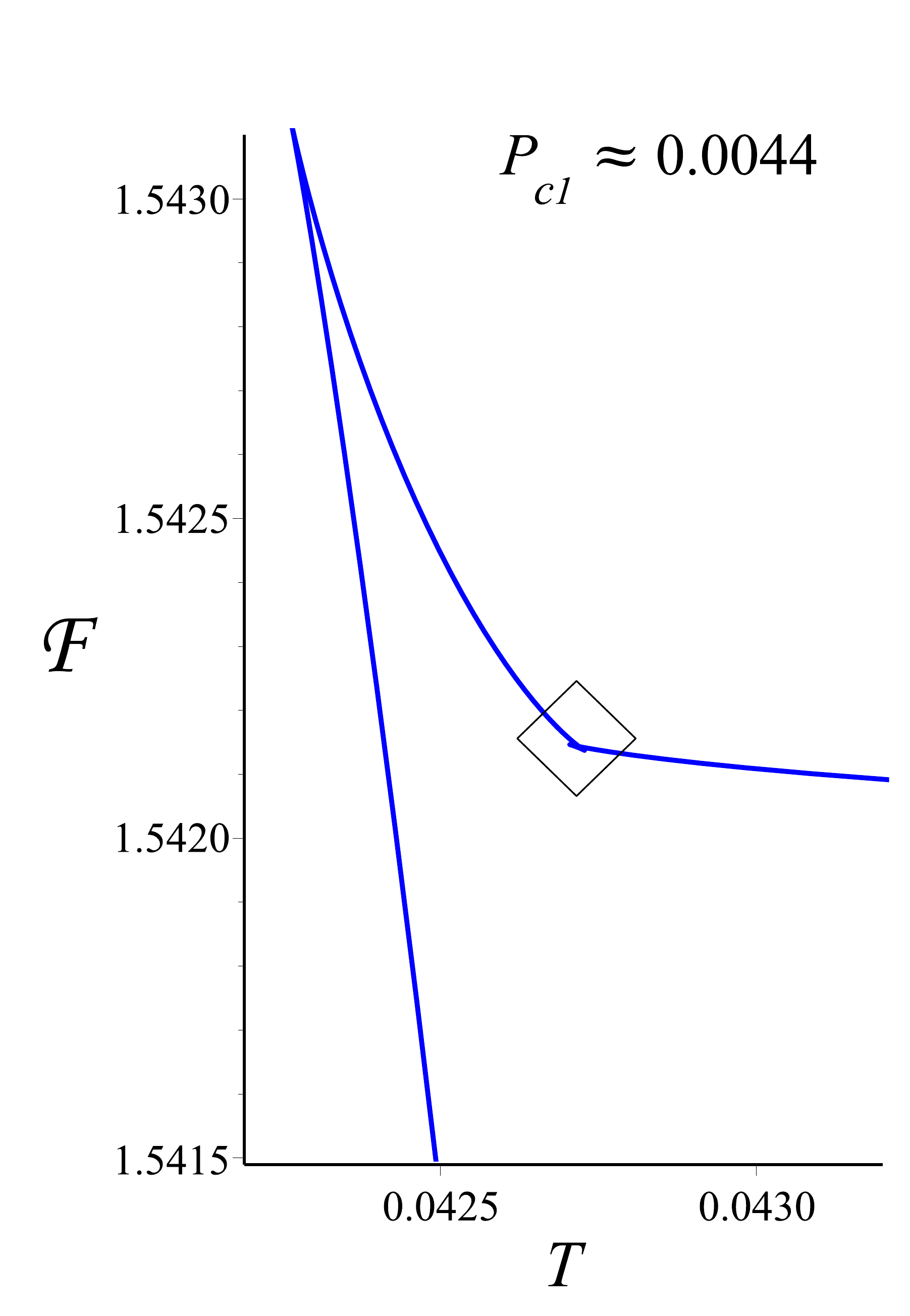}}
\subfigure[]
{\includegraphics[scale=0.17]{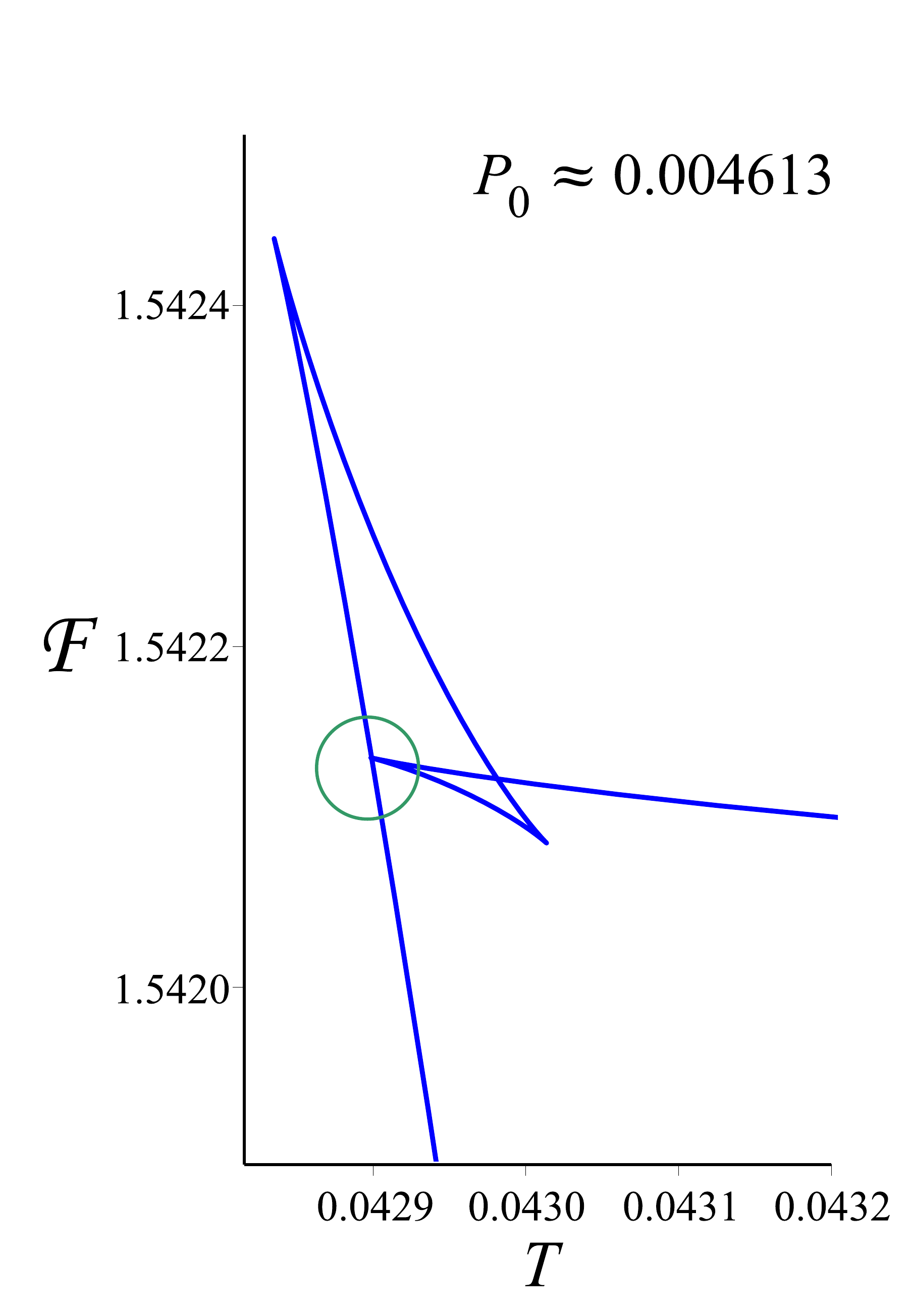}}
\subfigure[]
{\includegraphics[scale=0.17]{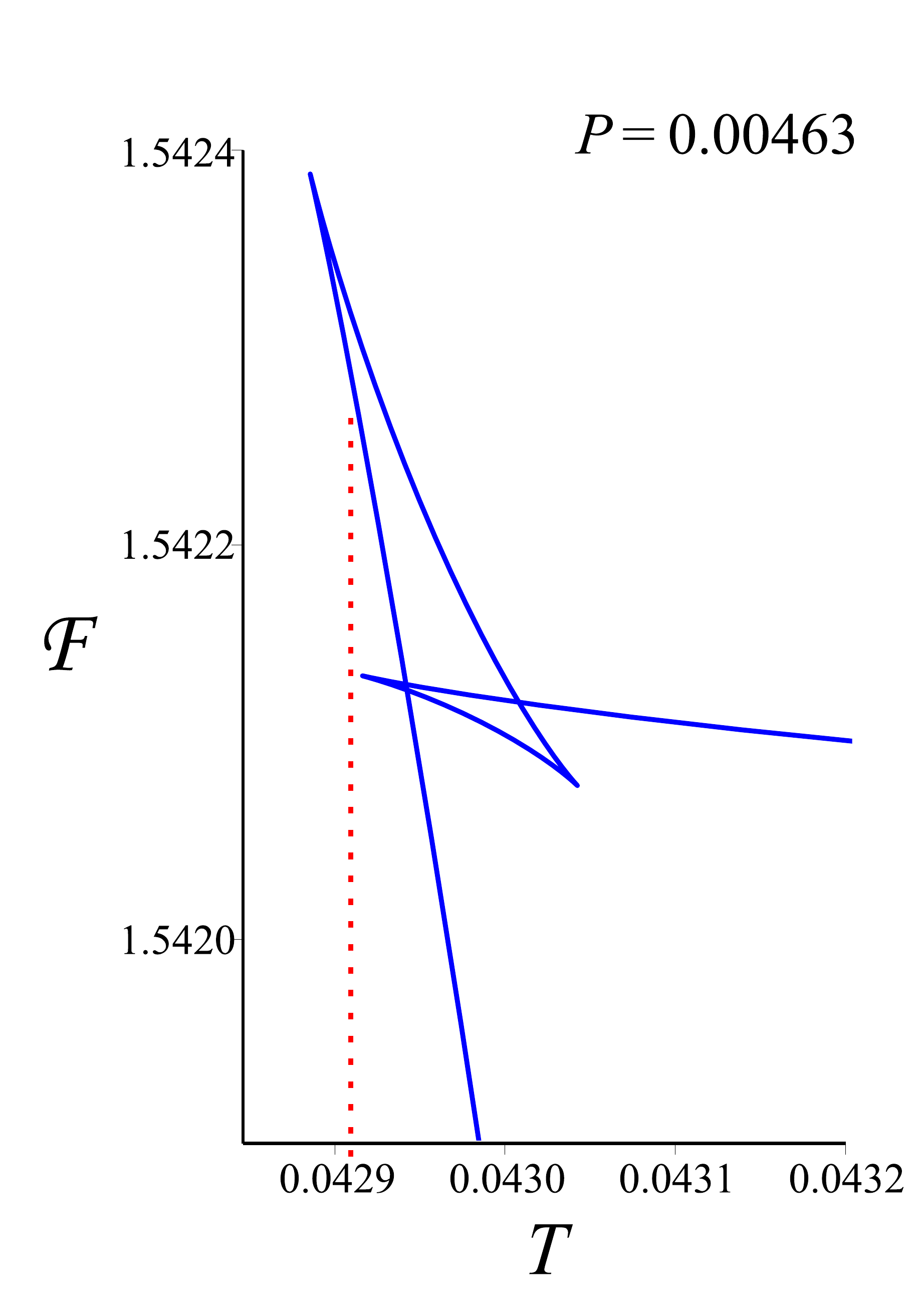}}
\subfigure[]
{\includegraphics[scale=0.17]{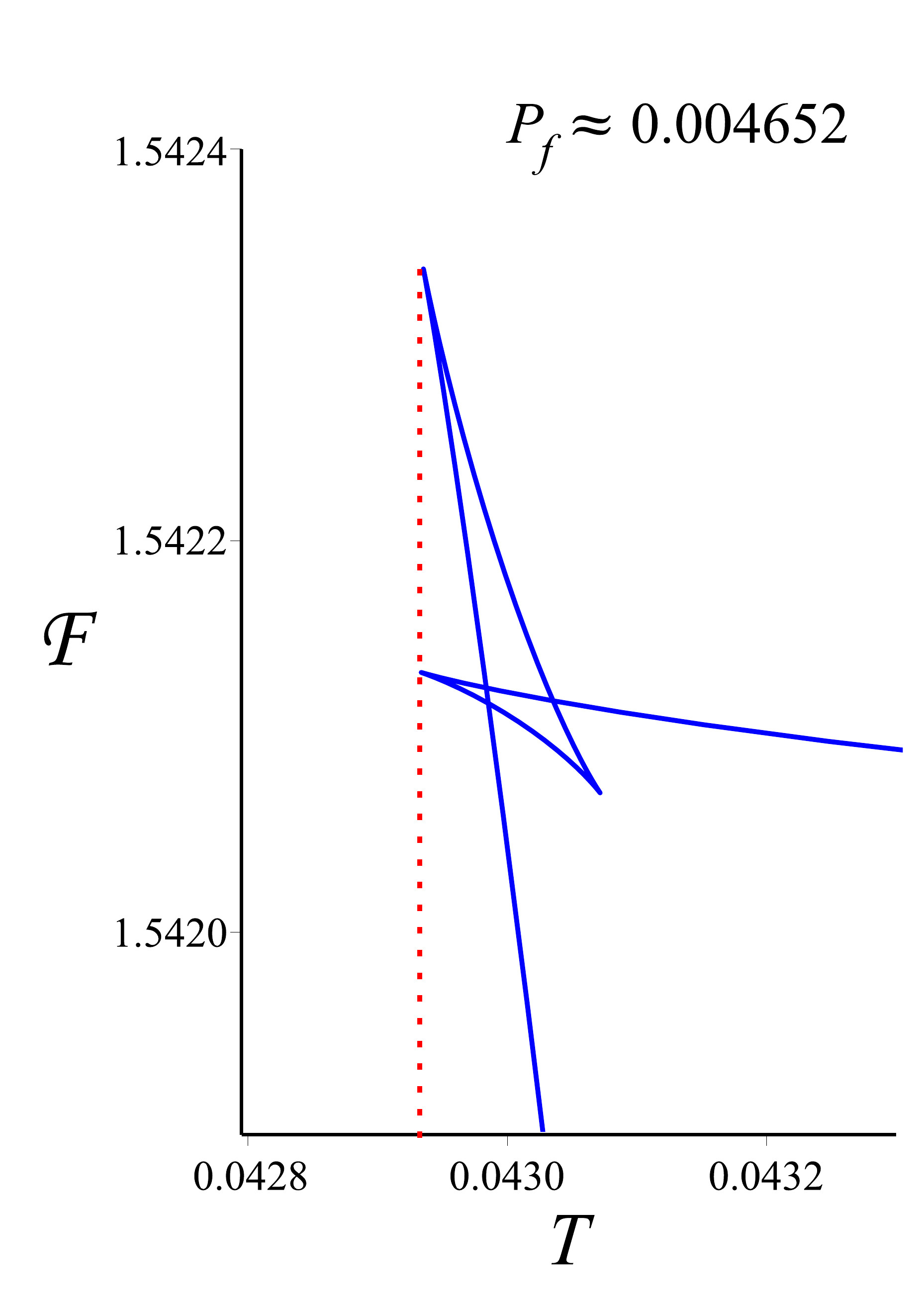}}
\subfigure[]
{\includegraphics[scale=0.17]{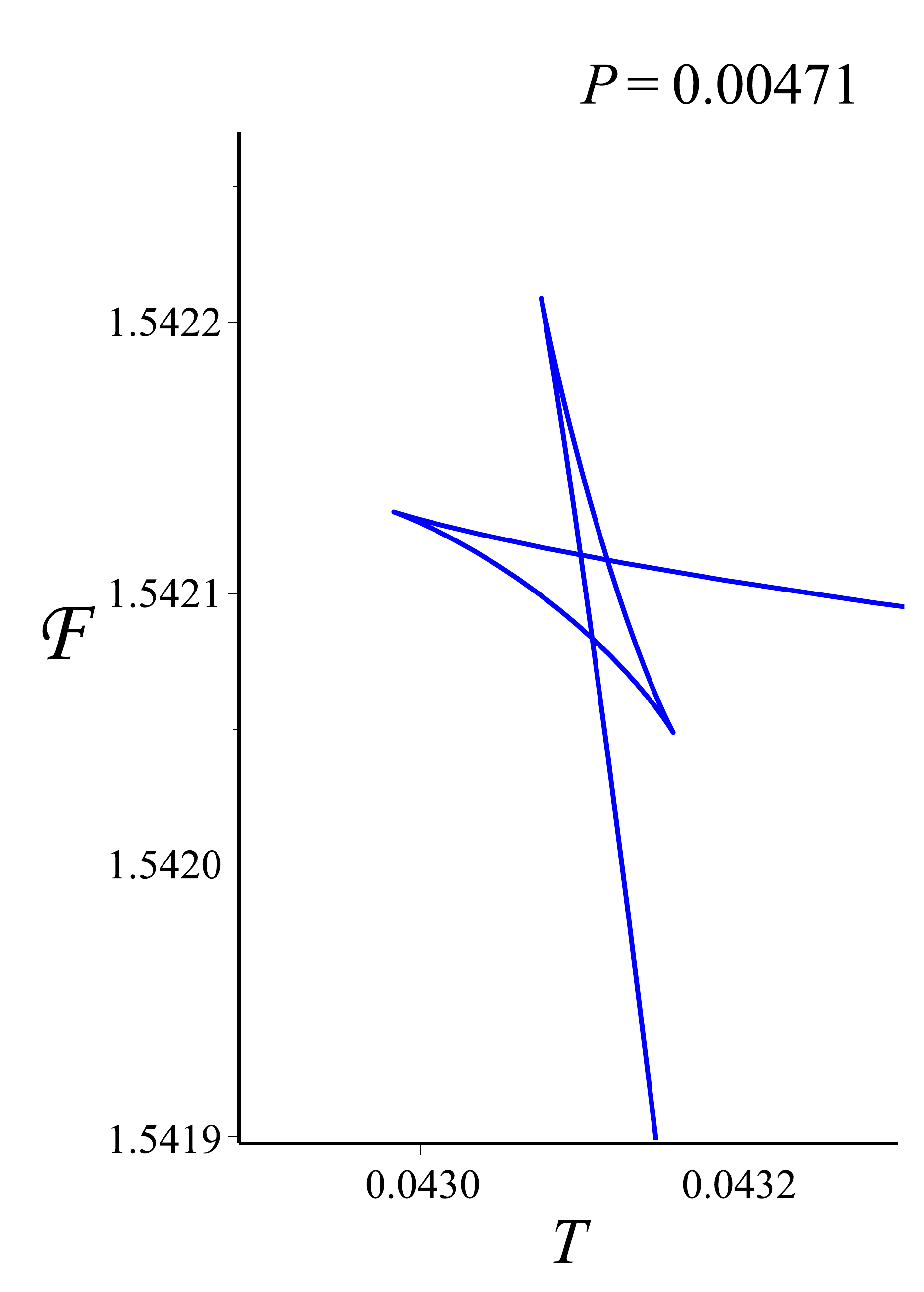}}
\subfigure[]
{\includegraphics[scale=0.17]{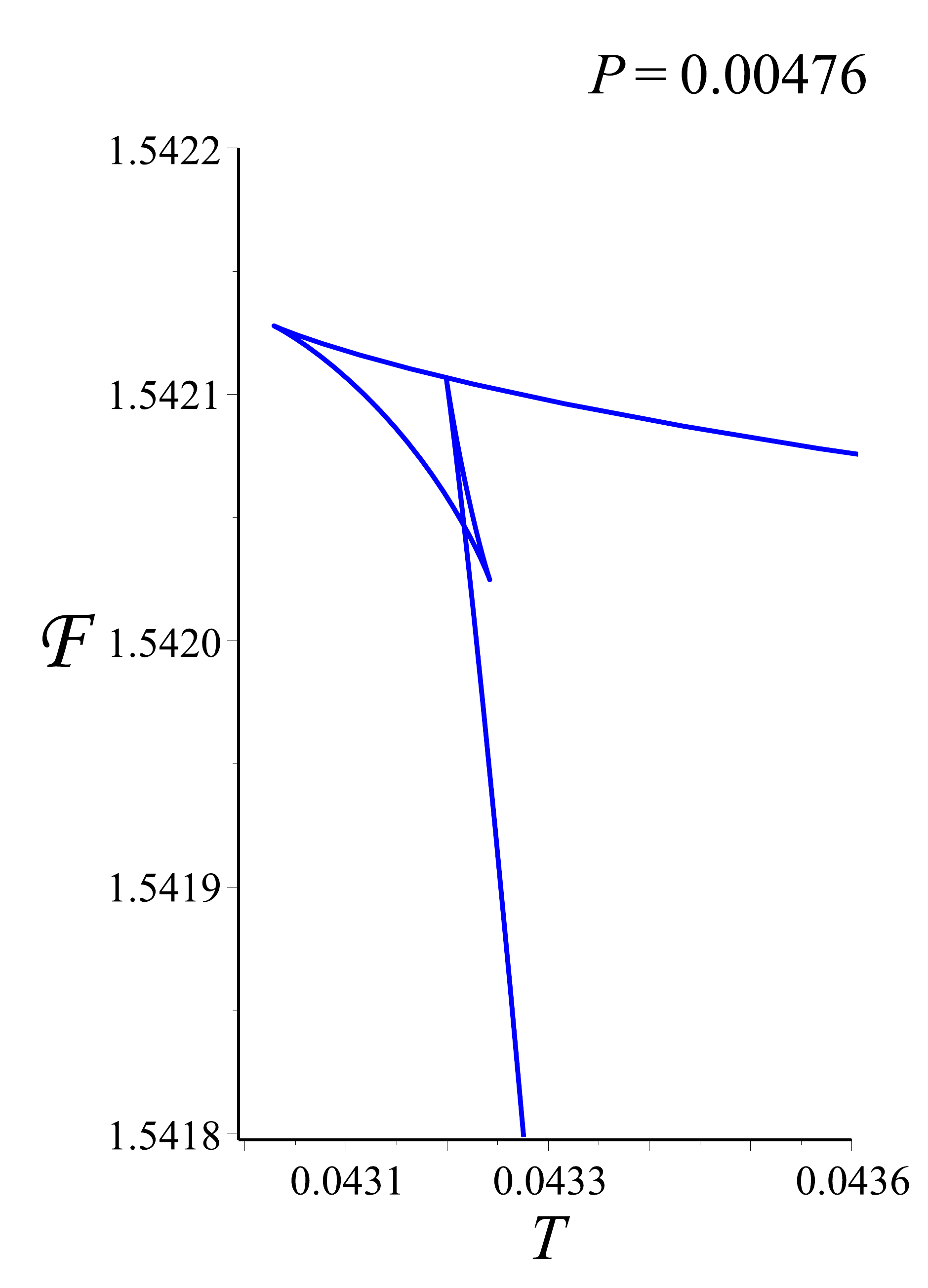}}
\subfigure[]
{\includegraphics[scale=0.17]{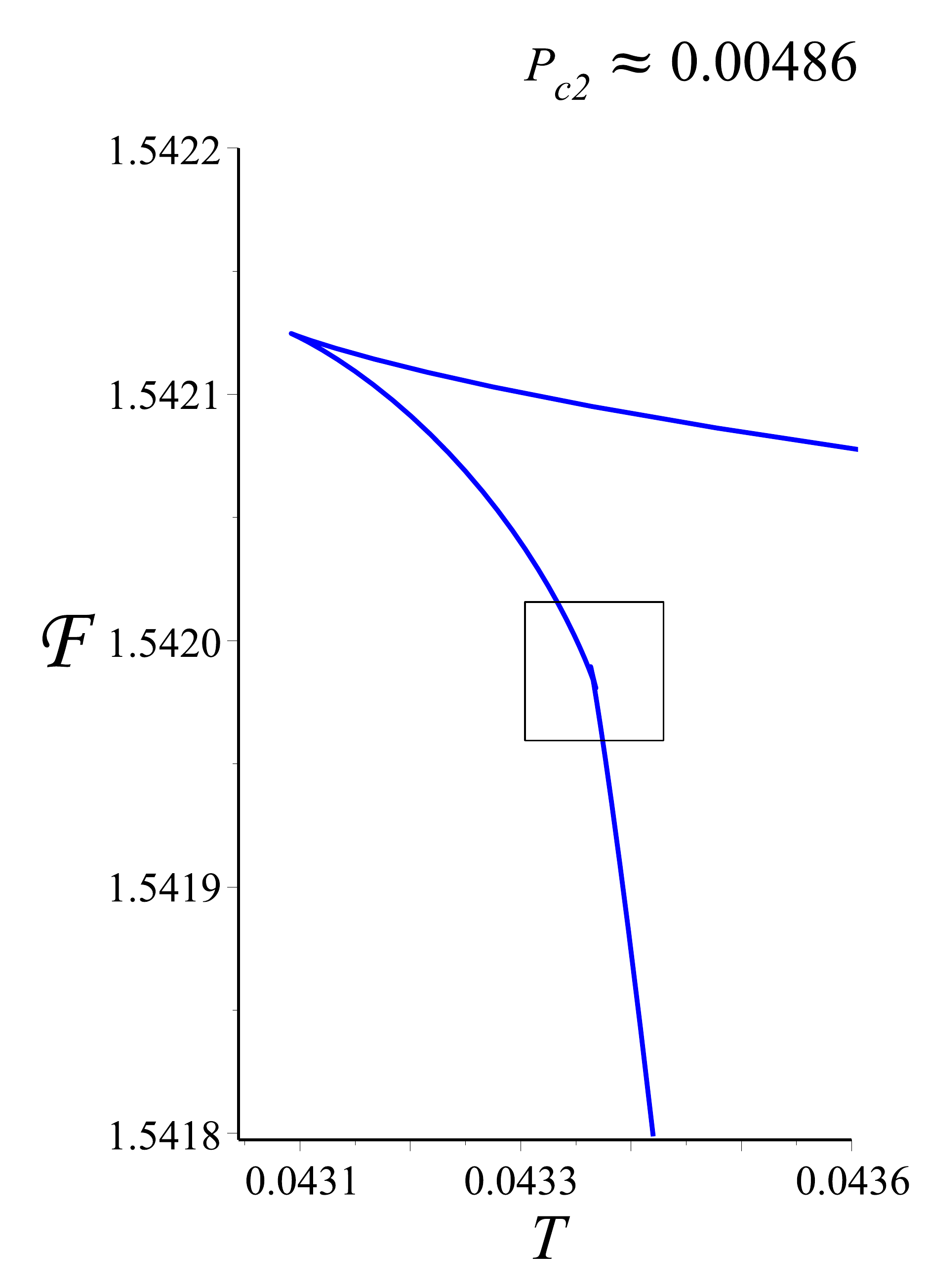}}
\subfigure[]
{\includegraphics[scale=0.17]{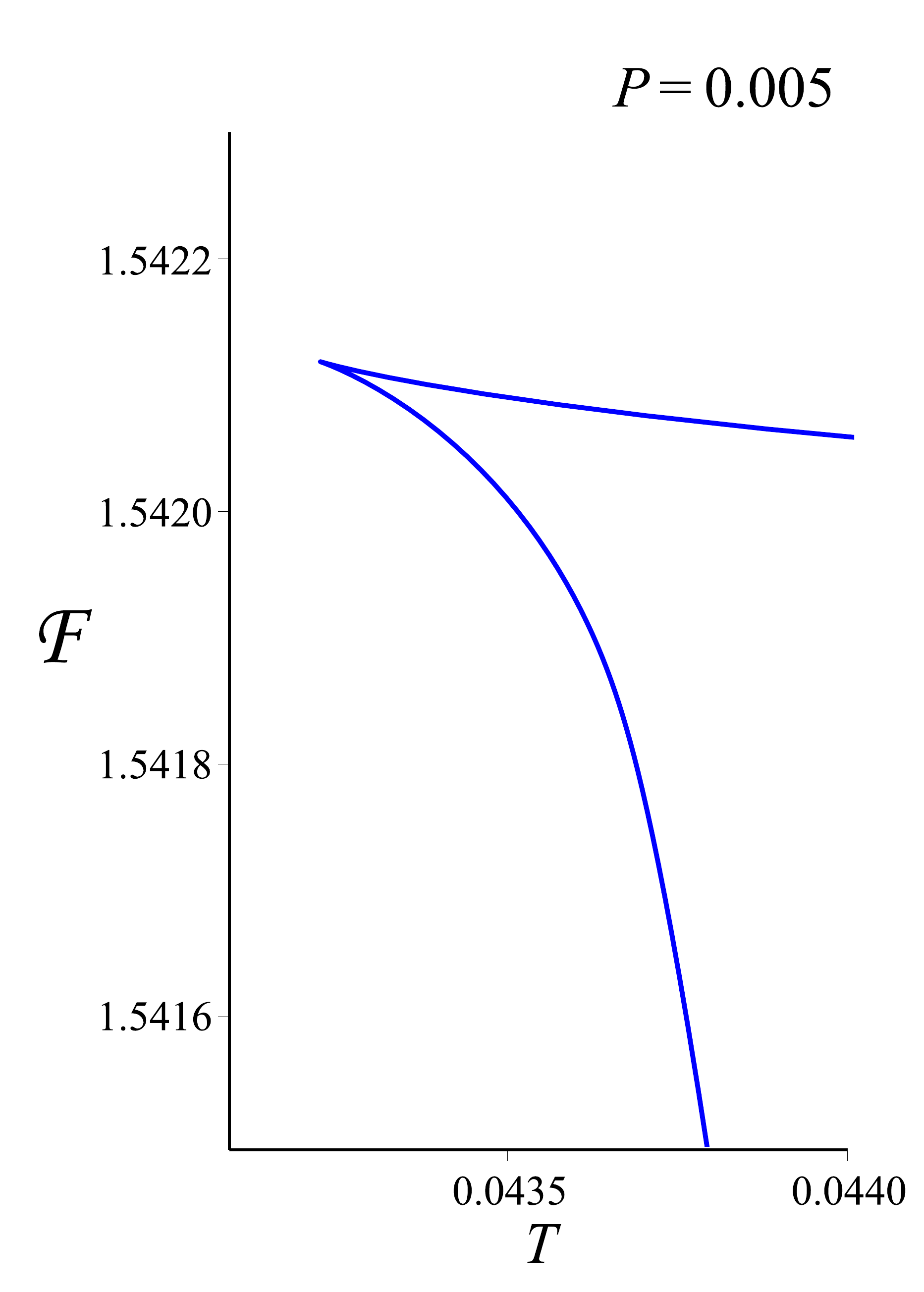}}
\caption{\small $\mathcal{F}$ vs $T$, $Q=2.65$, for increasing values of $P$.}
\label{FT0}
\end{figure}

Within the range of double criticality, $Q_{min}<Q<Q_0$, the situation is much more interesting. In Fig. \ref{FT0}, we show a number of snapshots for a relevant range of $P$ for $Q=2.65$.  At $P_{c1}$, the first critical point appears, after which an ``inverted" swallowtail emerges. This swallowtail does not signify a phase transition since all points on it are above the global minimum of the free energy.  
However, as the pressure further increases, the inverted swallowtail moves leftward and, eventually, at $P_0$ it intersects the lower part of the curve, i.e., the branch of large stable black holes, as shown in Fig. \ref{FT0}(b) with the intersection marked by a green circle. For larger pressures, we have the standard large-to-small first order phase transition that takes place until $P=P_{c2}$. In addition, we observe a zeroth order phase transition that consists of a ``jump" in the value of the thermodynamic potential, as shown by the red dotted line in Fig. \ref{FT0}(c). This phase transition starts from $P=P_0$ and extends until $P=P_f$, when the temperatures of the left parts of the swallowtails coincide, as shown in Fig. \ref{FT0}(d). Beyond this it moves further left, yielding a second swallowtail (the standard one). The upper cusp of the standard swallowtail moves downward as pressure further increases until it intersects the upper part of the inverted swallowtail, shown in Fig. \ref{FT0}(f),  after which the inverted one disappears. The standard one continues to shrink as pressure increases until, at the second critical point $P=P_{c2}\approx 0.0049$, the standard swallowtail also disappears, as shown in Fig. \ref{FT0}(g). There is no critical behavior for $P>P_{c2}$.

\begin{figure}[h!]
\centering
\includegraphics[scale=0.42]{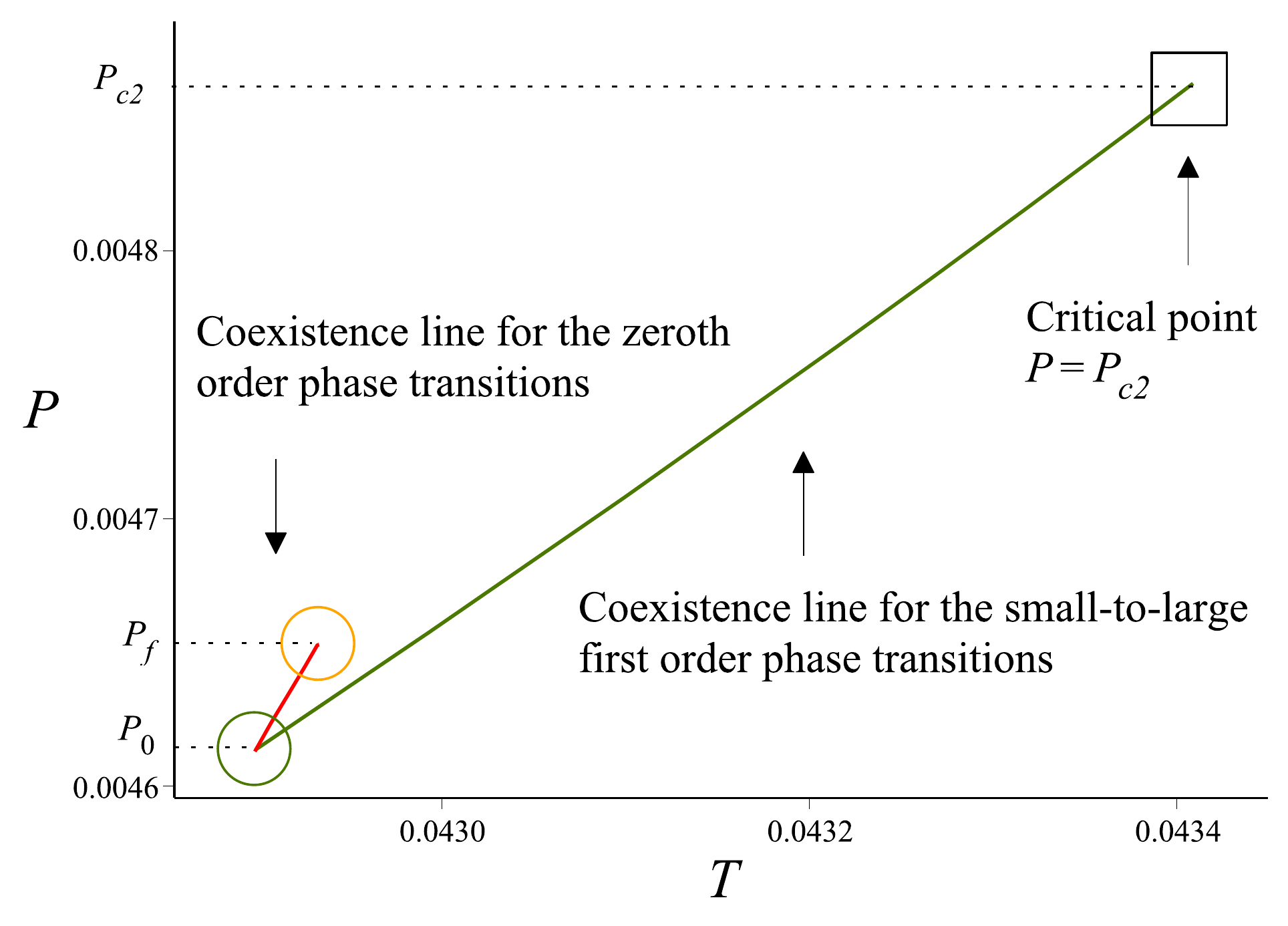}
\caption{\small Coexistence lines for zeroth order phase transitions and first order phase transitions. They meet at the large/small/large triple point at $P=P_0$.}
\label{coex}
\end{figure}

This situation is better illustrated in Fig. \ref{coex}, in which the coexistence lines for the corresponding phase transitions are depicted. We see that, for $P_f > P > P_0$, there are two coexistence lines and, within this range of pressures, we have reentrant phase transitions from large to small to large black holes as the temperature continuously decreases.

\noindent\emph{Discussion ---}
We have seen that the presence of scalar field considerably modifies the Van der Waals behavior previously observed for charged black holes \cite{Kubiznak:2012wp,Gunasekaran:2012dq}. A phase transition is said to be reentrant if it involves the transformation of a system from one state into a macroscopically similar state via at least two phase transitions through the variation of a single thermodynamic parameter \cite{Narayanan}.  By studying a class of charged hairy black holes that have finite horizon area in the zero temperature limit \cite{Anabalon:2013sra}, we have found that there is an intermediate range of the charge in which reentrant phase behavior is exhibited. A different class of hairy black holes was studied in Ref. \cite{Astefanesei:2019ehu}, but there are no reentrant phase transitions in this case. This is another interesting and concrete example on the analogy between black hole physics and condensed matter. 

The scalar field is ``secondary hair" in the sense that there is no associated integration constant and the scalar charge, which is not conserved, does not appear in the first law \cite{Astefanesei:2018vga}. Previously, we proved that the reverse isoperimetric inequality is satisfied. The fact that $\mathcal{R}\geq 1$ has the  following simple interpretation: for a black hole with a given thermodynamic volume, the charged hairy AdS black hole carries less entropy than the RN-AdS counterpart. This is expected because the remaining entropy is taken by the scalar field (``hair" degrees of freedom) beyond the event horizon.

The nontrivial thermodynamic behavior and the fact that there exists a well defined extremal limit are not due to the coupling between the electric and scalar fields, but rather to the special scalar potential that is coming from supergravity. The physical intuition behind the existence of a finite horizon area in the extremal limit is that there is a competition between the effective potential due to the coupling with the gauge field and the potential of the theory, which makes the extremal black hole regular. It is also important to emphasize that the critical behavior appears for a specific range of the parameters in the potential. If we restore the constant $\alpha$, we obtain that criticality is restricted to theories satisfying
\begin{equation}
-{0.134}\alpha\lessapprox\Lambda\leq 0
\end{equation}
in accordance with the second plot in Fig. \ref{PT}.

To close, we showed that the self-interaction of the scalar field can change drastically the thermodynamic behavior of black holes. These examples hint at the fact that in more realistic scenarios, e.g., if the dark matter is a scalar field
with self-interaction, the phase diagram is richer than that for black holes with no hair. Our study was done for theories with negative cosmological constant. Interestingly, in the limit $\Lambda\rightarrow0$, the critical temperature remains finite for $Q=Q_0$ (as in Figure \ref{PT}) that is a sign of criticality in asymptotically flat spacetime. We hope to report on this in the near future. 

\noindent\emph{Acknowledgments}
This work was supported in part by the Natural Sciences and Engineering Research Council of Canada. The work of D.A. and R.R. was supported by the Fondecyt Grant No. 1200986, and the work of P.C. was supported by CONICYT (currently ANID) Grant No. 21182145. 


\end{document}